\documentclass[12pt]{iopart}
\usepackage{units}
\usepackage{graphicx}
\usepackage{cite}
\usepackage{amssymb}

\usepackage{textcomp}

\begin{document}
\title[Anisotropy study of multiferroicity in the pyroxene NaFeGe$_2$O$_6$]{Anisotropy study of multiferroicity in the pyroxene NaFeGe$_2$O$_6$}

\author{M Ackermann$^{1}$, L Andersen$^{2}$, T Lorenz$^{2}$,
L Bohat\'y$^1$ and P Becker$^{1}$}

\address{$^1$ Institut f\"ur Kristallographie, Universit\"at zu K\"oln, Greinstra\ss e 6, 50939 K\"oln, Germany}
\address{$^2$ II. Physikalisches Institut, Universit\"at zu K\"oln, Z\"ulpicher Stra\ss e 77, 50937 K\"oln, Germany}

\ead{petra.becker@uni-koeln.de}

\begin{abstract}
We present a study of the anisotropy of the dielectric, magnetic and magnetoelastic properties of the multiferroic clinopyroxene NaFeGe$_2$O$_6$. Pyroelectric currents, dielectric constants and magnetic susceptibilities as well as the thermal expansion and the magnetostriction were examined on large synthetic single crystals of NaFeGe$_2$O$_6$. The spontaneous electric polarization detected below $T_{\rm C}\simeq\unit[11.6]{K}$ in an antiferromagnetically ordered state ($T_{\rm N}\simeq\unit[13]{K}$) is mainly lying within the $ac$ plane with a small component along $\bi{b}$, indicating a triclinic symmetry of the multiferroic phase of NaFeGe$_2$O$_6$. The electric polarization can be strongly modified by applying magnetic fields along different directions. We derive detailed magnetic-field versus temperature phase diagrams and identify three multiferroic low-temperature phases, which are separated by a non-ferroelectric, antiferromagnetically ordered state from the paramagnetic high-temperature phase.

\end{abstract}

\pacs{75.85.+t, 77.80.-e, 75.30.Cr, 77.70.+a}
\submitto{\NJP}
\maketitle

\section{Introduction}
\label{intro}
Spin-driven multiferroics with complex antiferromagnetic orders, such as spiral or cycloidal spin structures, are in the focus of scientific interest since the last ten years~\cite{Kimura2003, spaldin2005renaissance,cheong2007multiferroics,kimura2007spiral}.  Because in this class of materials ferroelectricity usually is induced by a primary magnetic ordering, ferroelectric and magnetic order show strong magnetoelectric coupling. Therefore, the rather small spontaneous electric polarizations, compared to conventional ferroelectrics, can be strongly modified and controlled by external magnetic fields. As a result of the intense research activities, a series of new spin-driven multiferroic materials was found in the last few years, e.g.~\cite{kimura2005magnetoelectric,Lawes2005,Park2007,Heyer,Taniguchi,Arkenbout, Kimura2008,Johnson2012,Seki2010,Blinc2008,Ackermann2013,Ackermann2014}.  
In particular the pyroxenes (general formula $AMX_2$O$_6$ with $A=$ mono-  or divalent metal, \mbox{$M=$ di-} or trivalent metal and $X=$ tri-  or tetravalent cation) form a huge family of potentially multiferroic and magnetoelectric materials~\cite{Jodlauk2007}. The mineral aegirine of the composition Na$_{1.04}$Fe$_{0.83}$Ca$_{0.04}$Mn$_{0.02}$Al$_{0.01}$Ti$_{0.08}$Si$_2$O$_6$ was identified as multiferroic a few years ago, while LiFeSi$_2$O$_6$ and LiCrSi$_2$O$_6$ were found to be linear magnetoelectrics~\cite{Jodlauk2007}. Interestingly, synthetic NaFeSi$_2$O$_6$ apparently has a modified magnetic structure compared to natural aegirine indicating that it probably is only linear magnetoelectric~\cite{Baker2010}. Although in the years after the pioneering work of Jodlauk {\it et al.}~\cite{Jodlauk2007} intense research activities set in to find more multiferroic materials within the pyroxene family~\mbox{\cite{Redhammer2009, Nenert2009, Nenert2010, Nenert2009b, Redhammer2011}}, up to date only one further multiferroic compound could be identified, namely NaFeGe$_2$O$_6$~\cite{Kim2012}.

\begin{figure}[t]
\includegraphics[width=\textwidth]{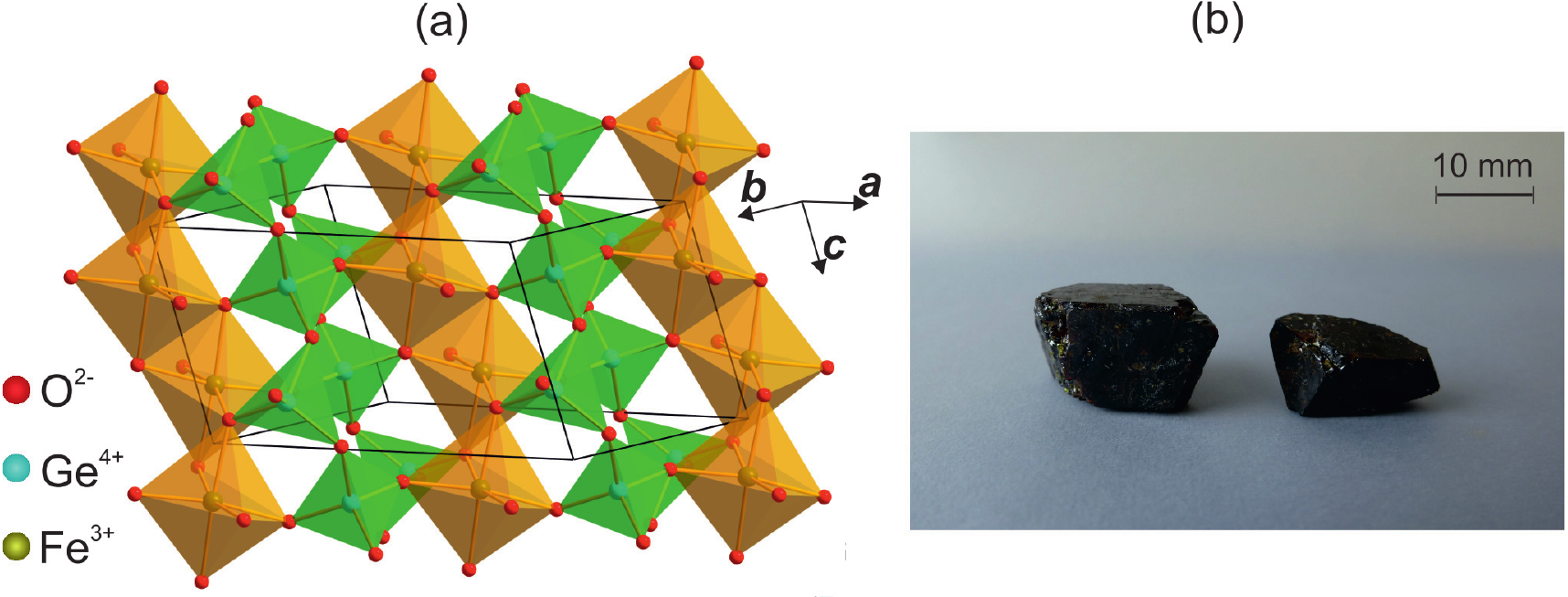}
\caption{(a) Main features of the crystal structure of NaFeGe$_2$O$_6$. The chains of edge-sharing [FeO$_6$] octahedra (orange) running along $\bi{c}$ are linked via chains of corner sharing [GeO$_4$] tetrahedra (green). The structural data are taken from~\cite{Redhammer2011}. (b) Synthetic crystals of NaFeGe$_2$O$_6$.}
\label{figure0}
\end{figure}

NaFeGe$_2$O$_6$ belongs to the subgroup of clinopyroxenes and crystallizes at room temperature in the space group $C2/c$ with the lattice parameters $a=\unit[10.0073(8)]{\AA}$, $b=\unit[8.9382(7)]{\AA}$, $c=\unit[5.5184(4)]{\AA}$ and $\beta=\unit[107.524(1)]{^{\circ}}$~\cite{Redhammer2011}. The structure consists of one-dimensional zigzag chains of edge-sharing FeO$_6$ octahedra, which are connected by chains of  GeO$_4$ tetrahedra within the $(110)$ and $(\bar{1}10)$ planes. Both chain systems are running along the $\bi{c}$ axis, see figure~\ref{figure0}\,(a). There are three relevant magnetic exchange interactions $J_1, J_2$ and $J_3$~\cite{Streltsov}. Along the zigzag chains $J_1$ connects neighbouring Fe$^{3+}$ sites via Fe--O--Fe super-exchange pathways. The super-exchange interactions $J_2$ and $J_3$ connect Fe$^{3+}$ sites of different chains via one or two [GeO$_4$] tetrahedra, respectively. In this context, the Fe$^{3+}$ moments form triangular lattices within the $(110)$ and $(\bar{1}10)$ planes, which can give rise to a magnetic frustration.

The vast majority of the investigations of NaFeGe$_2$O$_6$ were performed on polycrystalline powder samples~\cite{Drokina2008, Drokina2009,Drokina2011,Redhammer2011,Kim2012}. Measurements of the magnetic susceptibility revealed the occurrence of low-dimensional magnetic correlations around \unit[35]{K}, succeeded by the onset of a three-dimensional antiferromagnetic order below $\simeq\unit[13]{K}$~\cite{Drokina2011,Redhammer2011,Kim2012}. Previous results of magnetic-susceptibility and  M\"ossbauer-spectroscopy measurements had indicated a slightly higher N\'eel temperature of about \unit[15]{K}~\cite{Drokina2008, Drokina2009}. Among the group of Fe$^{3+}$-based pyroxenes, NaFeGe$_2$O$_6$ exhibits the most pronounced low-dimensional magnetic characteristics~\cite{Redhammer2011}. Specific-heat measurements revealed a further phase transition at about \unit[12]{K}~\cite{Drokina2011}. Dielectric investigations on sintered polycrystalline pellets revealed that this second transition coincides with the onset of a spontaneous polarization of about $\unit[13]{\mu C/m^2}$, which decreases with increasing magnetic field~\cite{Kim2012}. Two different neutron-diffraction experiments on powder as well as on single-crystal samples of NaFeGe$_2$O$_6$ are reported in literature~\cite{Drokina2010, Redhammer2011, Drokina2011}. The results of both studies indicate that the magnetic structure of NaFeGe$_2$O$_6$ forms an incommensurate cycloidal spin arrangement below the second transition at $\simeq\unit[12]{K}$ with the spins lying mainly within the $ac$ plane. The reported propagation vectors $\bi{k}=(0.3357,0,0.0814)$~\cite{Drokina2010,Drokina2011} and $\bi{k}'=(0.323,1.0,0.080)$~\cite{Redhammer2011}, however, are contradictory and the magnetic structure between $\simeq\unit[12]{K}$ and $\simeq\unit[13]{K}$ has not been resolved yet.

Here, we present a detailed study of thermodynamic properties of NaFeGe$_2$O$_6$, using large synthetic single crystals, elucidating the whole anisotropy of its magnetic and multiferroic properties. The spontaneous electric polarization detected below $T_{\rm C}\simeq\unit[11.6]{K}$ is mainly lying within the $ac$ plane with a small component along $\bi{b}$, indicating a triclinic symmetry of the multiferroic phase of NaFeGe$_2$O$_6$. The electric polarization can strongly be modified by applying magnetic fields, which induce transitions to other phases. The paper is organized as follows. First, the crystal growth of NaFeGe$_2$O$_6$ and the experimental techniques for the study of its multiferroic properties are described. Then, the results of the magnetic-susceptibility measurements and the dielectric investigations are presented and discussed.  Combining these data with measurements of thermal expansion and magnetostriction, detailed magnetic-field versus temperature phase diagrams are derived. Finally the multiferroic properties of NaFeGe$_2$O$_6$ are compared with those of the first multiferroic pyroxene, the mineral aegirine (of the composition Na$_{1.04}$Fe$_{0.83}$Ca$_{0.04}$Mn$_{0.02}$Al$_{0.01}$Ti$_{0.08}$Si$_2$O$_6$).

\section{Experiments}
\label{sec:1}

Due to their incongruent melting behaviour single crystals of NaFeGe$_2$O$_6$ were grown from high-temperature solution by the top seeding technique. By the use of a solvent of nearly eutectic composition from the system Na$_2$MoO$_4$--NaVO$_3$ and a ratio solvent\,:\,NaFeGe$_2$O$_6$ of 4\,:\,1, untwinned large single crystals of high quality were obtained and the formation of additional parasitic phases, such as hematite ($\alpha$-Fe$_2$O$_3$) and maghemite ($\gamma$-Fe$_2$O$_3$) could be suppressed to a large extend. The growth was performed in the temperature range between \unit[1295]{K} and \unit[1288]{K} with an applied cooling rate of \unit[0.2]{K/day}, which resulted in crystals of dimensions up to $\unit[10\times 10\times 15]{mm^3}$ with a well-developed morphology, see figure~\ref{figure0}\,(b). The crystal morphology is dominated by the prisms $\{110\}$ and $\{\bar{1}11\}$ and, minor, the pinacoids $\{100\}$ and $\{010\}$. The morphological faces, which were identified via X-ray diffraction, were used as reference planes for the sample orientation. Samples with faces perpendicular to $\bi{b}\times\bi{c}$, $\bi{b}$ and $\bi{c}$ were prepared. The magnetic susceptibility, thermal expansion and magnetostriction were measured on one single sample with dimensions of $\simeq 2\times 1\times 1$~mm$^3$.  The dielectric investigations were performed on typically \unit[1]{mm} thick plates of surfaces in the range of $\sim$\,$\unit[30]{mm^2}$, which were vapour-metallized with silver electrodes. 

The magnetization was measured with a commercial vibrating sample magnetometer (PPMS, Quantum Design)  from about \unit[2]{K} to room temperature in magnetic fields up to \unit[14]{T}. The dielectric measurements were performed in the temperature range from about 3 to \unit[25]{K} in a cryostat equipped with a \unit[15]{T} magnet and a variable temperature insert (KONTI cryostat, CryoVac). The electric polarization was calculated via time integration of the pyroelectric currents measured by an electrometer (Keithley 6517) as a function of increasing temperature. During the cooling process, an electric poling field of at least \unit[200]{V/mm} was applied well above the ordering transition temperatures in order to reach a single-domain phase. The poling fields were removed at base temperature and the pyroelectric currents was recorded while heating the sample with a rate of \unit[3]{K/min}. In all cases, the electric polarization could be completely inverted  by reversing the electric poling field. In the same setup we also determined the relative dielectric constants $\epsilon_{i}^{r}$ ($i = b\times c, b, c$)\footnote{Tensors are related to a Cartesian reference system with unit axes running along the directions of $\bi{b}\times\bi{c}$, $\bi{b}$ and $\bi{c}$, where $\bi{b}$ and $\bi{c}$ are the crystallographic axes.} from the measured capacitance of the metallized samples as a function of temperature or magnetic field using a capacitance bridge (Andeen-Hagerling 2500A) at a frequency of \unit[1]{kHz}. The thermal expansion and magnetostriction up to  \unit[15]{T} were measured on a home-built capacitance dilatometer in the temperature range between 3 and \unit[15]{K}. The magnetic field was applied either along $\bi{b}\times\bi{c}$, $\bi{b}$, or $\bi{c}$ and in each case the length change $\Delta L_c(T,\bi{B})$ along the $\bi{c}$ axis was measured, either as a function of continuously varying $T$ or $\bi{B}$ with rates of $\pm 0.05$ to $\pm \unit[0.1]{K/min}$ or $\pm \unit[0.1]{T/min}$, respectively. The thermal-expansion coefficient $\alpha_c = 1/L_c^0 \cdot \partial  \Delta L_c/\partial T$ was obtained by numerically calculating the temperature derivative of the length change $\Delta L_c$, where $L_c^0$ denotes the sample length along $\bi{c}$.

\section{Results and discussion}

\begin{figure}[t]
\includegraphics[width=\textwidth]{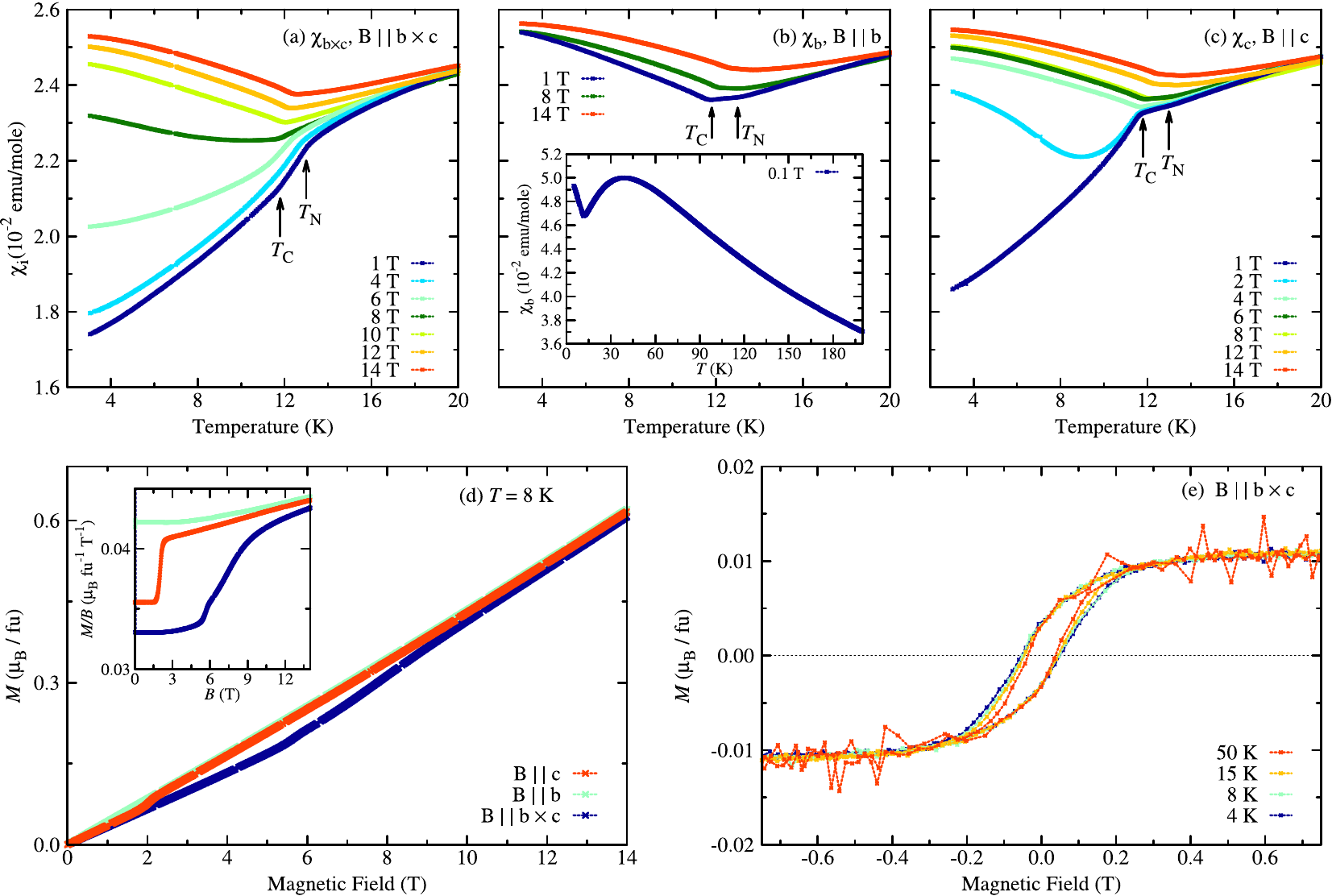}
\caption{(a)-(c) Magnetic susceptibilities $\chi_{b \times c}$, $\chi_{b}$, and $\chi_{c}$ of NaFeGe$_2$O$_6$ for different magnetic fields as functions of temperature. The inset in (b) shows $\chi_{b}$, measured at \unit[0.1]{T} in a wider temperature range. (d) Magnetic-field dependent magnetization at \unit[8]{K} for different field directions; the inset displays the same data as $M/B$ versus $B$. (e) Hysteretic magnetization exemplified for $\bi{B}||\bi{b}\!\!\times\!\!\bi{c}$ at different temperatures, which probably arises from an impurity phase of ferrimagnetic $\gamma$-Fe$_2$O$_3$.}
\label{figure1}
\end{figure}

Figure~\ref{figure1} displays measurements of the magnetic susceptibility and the magnetization of NaFeGe$_2$O$_6$.  Magnetic fields up to 14~T were applied along the \mbox{$\bi{b}\!\!\times\!\!\bi{c}$}, $\bi{b}$ and $\bi{c}$ direction. The low-field curves confirm previous results~\cite{Drokina2011,Redhammer2011,Kim2012} and signal the onset of magnetic ordering at $T_{\rm N} \simeq \unit[13]{K}$ and a second transition at $T_{\rm C} \simeq \unit[11.6]{K}$, where apparently a spin reorientation occurs. Both, $\chi_{b}(T)$ and $\chi_{c}(T)$ have only weak anomalies at $T_{\rm N}$ and then hardly change in the temperature interval $[T_{\rm C},T_{\rm N}]$, while $\chi_{b\times c}(T)$ shows a pronounced kink at $T_{\rm N}$ and a subsequent decrease on lowering the temperature to $T_{\rm N}$. This indicates that the spins are oriented along the \mbox{$\bi{b}\!\!\times\!\!\bi{c}$} axis, which thus forms a magnetic easy axis below $T_{\rm N}$. The second transition at $T_{\rm C} \simeq \unit[11.6]{K}$ causes further kinks in all three susceptibilities. Below $T_{\rm C}$, $\chi_{b}(T)$ slightly increases with decreasing temperature whereas $\chi_{b\times c}(T)$ and $\chi_{c}(T)$ both decrease. This indicates that now the spins are within the $ac$ plane. The very similar behavior of $\chi_{b\times c}$ and $\chi_{c}$ and the fact that none of them approaches zero for  $T\rightarrow \unit[0]{K}$ implies, that there is no magnetic easy axis within the $ac$ plane below $T_{\rm C}$. Thus, in the low-temperature regime the spins of NaFeGe$_2$O$_6$ show an XY anisotropy with the $ac$ plane as magnetic easy plane. This result is in accordance with two different neutron diffraction studies~\cite{Drokina2010, Redhammer2011, Drokina2011}, which revealed that below \unit[5]{K} the spin structure of NaFeGe$_2$O$_6$ forms an incommensurate cycloid within the $ac$ plane. The magnetic structure between \unit[11.6]{K} and \unit[13]{K} has not been resolved yet. The inset of figure~\ref{figure1}\,(b) presents $\chi_{b}(T)$ up to \unit[200]{K}. At about \unit[35]{K} a broad maximum is visible confirming that the antiferromagnetic ordering in NaFeGe$_2$O$_6$ at $T_{\rm N} \simeq \unit[13]{K}$ is preceded by low-dimensional magnetic correlations, as already reported in literature~\cite{Redhammer2011}.

As can be seen in figure~\ref{figure1}\,(a) and (c), the decrease for $T<T_{\rm C}$ of both, $\chi_{b\times c}(T)$ and $\chi_{c}(T)$, systematically vanishes for larger fields and above about \unit[9]{T} the temperature dependences of $\chi_{i}(T)$ are almost identical for all three field directions. This is naturally explained by spin-flop transitions for \mbox{$\bi{B}|| \bi{b}\!\!\times\!\!\bi{c}$} or  $\bi{B}|| \bi{c}$ where the spin orientation changes from the easy $ac$ plane to the plane perpendicular to the respective magnetic field direction. As shown figure~\ref{figure1}\,(d), the spin-flop transitions cause abrupt changes in the corresponding low-temperature magnetization as a function of the magnetic field. The spin-flop fields at $T=\unit[8]{K}$ are $\simeq \unit[2]{T}$ and $\simeq \unit[5]{T}$ for \mbox{$\bi{B}|| \bi{b}\!\!\times\!\!\bi{c}$} and  $\bi{B}|| \bi{c}$, respectively. Indications for spin-flop transitions were already found in magnetic and dielectric investigations on polycrystalline samples of NaFeGe$_2$O$_6$~\cite{Kim2012}.  As will be shown below, structural changes and reorientations of the electric polarization accompany these spin-flop transitions. For $\bi{B}|| \bi{c}$, the transition is much sharper and the transition field is smaller than for \mbox{$\bi{B}|| \bi{b}\!\!\times\!\!\bi{c}$} illustrating that the magnetic properties of NaFeGe$_2$O$_6$ are not fully isotropic within the easy $ac$ plane. 

Around \unit[0]{T} the magnetization $M(B)$ shows a hysteretic behaviour for all magnetic field directions, which is exemplary illustrated in figure~\ref{figure1}\,(e) for \mbox{$\bi{B}|| \bi{b}\!\!\times\!\!\bi{c}$} at different temperatures. It is present in a wide temperature range also well above the magnetic ordering temperature of NaFeGe$_2$O$_6$ and the width of the hysteresis is nearly temperature independent. This indicates that the flux-grown NaFeGe$_2$O$_6$ crystals contain an impurity phase, probably maghemite $\gamma$-Fe$_2$O$_3$, which is ferrimagnetic and occasionally was formed during the crystal growth experiments, see section~\ref{sec:1}. From the observed saturation magnetization value of $\simeq\unit[0.01]{\mu_B}$ a contamination of less than 1\% can be estimated. The magnetic-susceptibility and magnetization data shown in figure~\ref{figure1}\,(a)-(d) have been corrected for the ferrimagnetic background signal.

\begin{figure}[t]
\includegraphics[width=\textwidth]{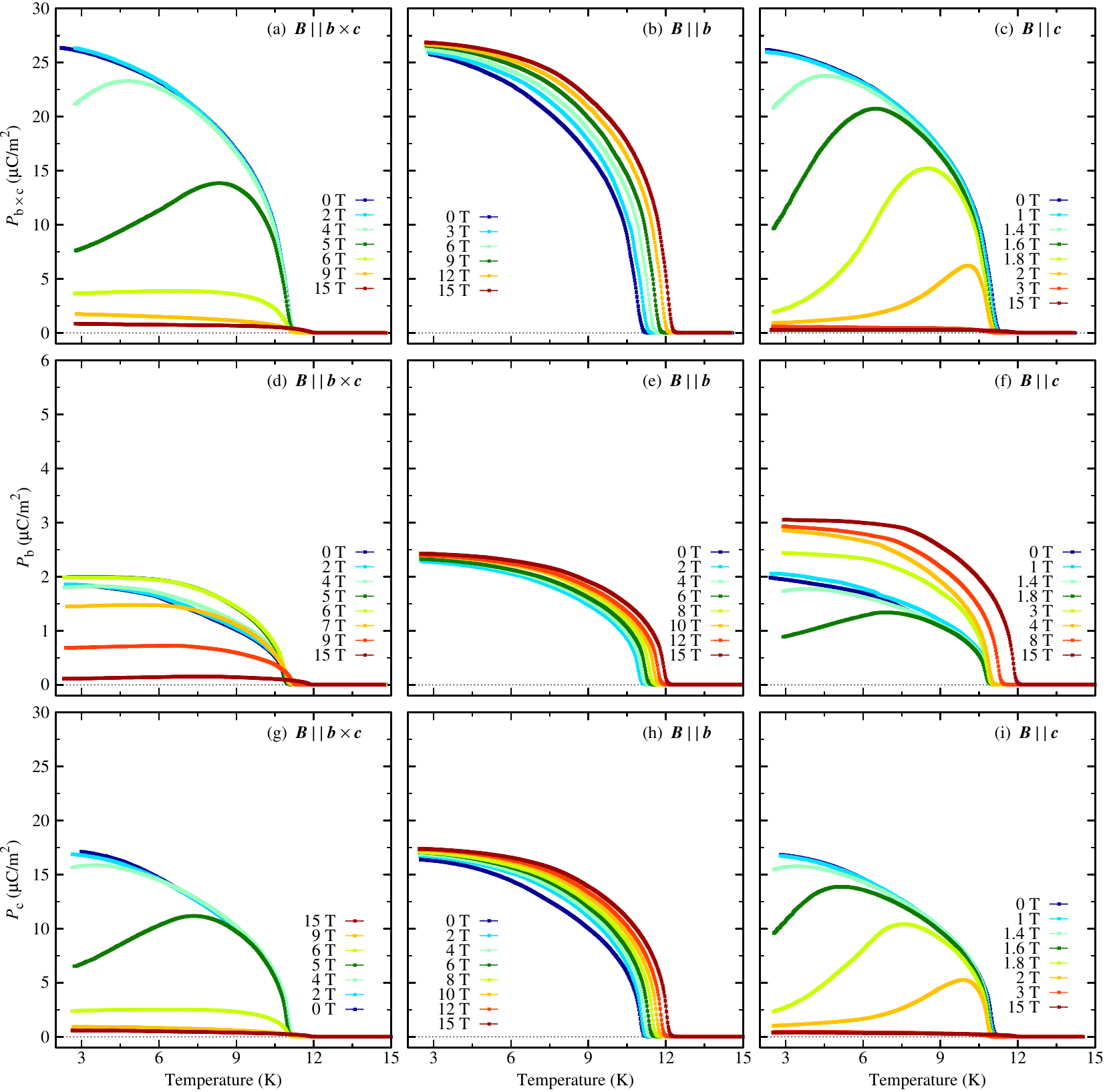}
\caption{Electric polarization $P_{b\times c}$, $P_b$ and $P_c$ (top to bottom) of NaFeGe$_2$O$_6$ measured as a function of temperature for constant magnetic fields applied parallel to the \mbox{$\bi{b}\!\!\times\!\!\bi{c}$}, $\bi{b}$ or $\bi{c}$ axis (left to right).}
\label{figure2}
\end{figure}

The electric polarization of NaFeGe$_2$O$_6$ depending on temperature and magnetic field is summarized in Figure~\ref{figure2}, which displays the components $P_{b\times c}$, $P_{b}$ and $P_{c}$ for magnetic fields applied either along \mbox{$\bi{b}\!\!\times\!\!\bi{c}$}, $\bi{b}$ or $\bi{c}$. As mentioned above, the electric polarization is completely invertible by inverting the electric poling fields. In figure~\ref{figure2}, only one poling direction is displayed. In zero magnetic field, a spontaneous electric polarization arises below $T_{\mathrm{C}}\simeq \unit[11.6]{K}$, with the components $P_{b\times c}\simeq \unit[27]{\mu C/m^{2}}$, $P_{b}\simeq \unit[2]{\mu C/m^{2}}$ and $P_{c}\simeq \unit[17]{\mu C/m^{2}}$ at \unit[3]{K}. This yields an absolute value of $\simeq \unit[32]{\mu C/m^{2}}$, which is almost lying within the $ac$ plane with a small component along $\bi{b}$. It is about a factor of 2.5 larger than the spontaneous electric polarization of aegirine (Na$_{1.04}$Fe$_{0.83}$Ca$_{0.04}$Mn$_{0.02}$Al$_{0.01}$Ti$_{0.08}$Si$_2$O$_6$), the second known multiferroic compound within the pyroxenes~\cite{Jodlauk2007}.  According to these results the crystallographic point group symmetry of the considered ferroelectric phase of  NaFeGe$_2$O$_6$ would be triclinic, 1.

\begin{figure}[t]
\includegraphics[width=\textwidth]{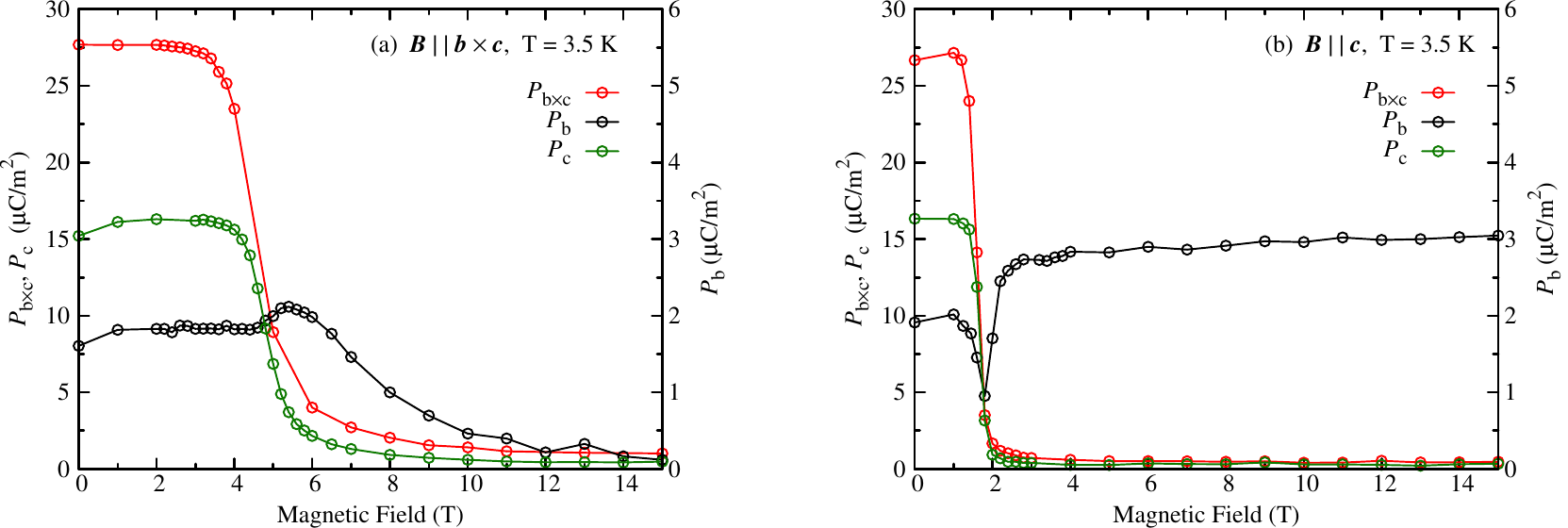}
\caption{Electric polarization $P_{b\times c}$, $P_b$ and $P_c$ as a function of magnetic field applied parallel to \mbox{$\bi{b}\!\!\times\!\!\bi{c}$} (left) or $\bi{c}$ (right) at $T=\unit[3.5]{K}$. The data are taken from the temperature-dependent polarization measurements shown in figure~\ref{figure2}. Note the different scales used for $P_{i}$ ($i=b\times c,c$) and for $P_b$.}
\label{figure2b}
\end{figure}

The electric polarization of many multiferroic materials with cycloidal spin structures can be described by the relation $\bi{P}\propto\bi{e_{ij}}\times(\bi{S^{(i)}}\times\bi{S^{(j)}})$, where $\bi{e_{ij}}$ denotes the unit vector connecting the ions with the spins $\bi{S^{(i)}}$ and $\bi{S^{(j)}}$~\cite{Sergienko2006, Katsura2005}. Taking into account the results of the magnetic investigations of the present work and those reported in literature with the spins lying within the $ac$ plane~\cite{Drokina2010, Redhammer2011, Drokina2011}, this relation would predict an electric polarization confined to the $ac$ plane as well, which is inconsistent with the present results. Either the spins in the cycloidal phase of NaFeGe$_2$O$_6$ need to have a finite component along $\bi{b}$, which was at least not excluded in~\cite{Drokina2011} or the model cannot be applied in this case. Another possible explanation for the observed inconsistency would be an erroneous sample orientation of the (010) sample. The observed polarization of $\simeq \unit[2]{\mu C/m^{2}}$ along $\bi{b}$ would require a misorientation of $\simeq\unit[4]{^{\circ}}$ from (010). This is, however, unlikely because investigations with a Laue camera revealed a maximum misorientation of less than $\unit[1.5]{^{\circ}}$. Moreover, the electric polarization along $\bi{b}$ was examined on two (010) samples cut from two different single crystals, which revealed absolute values for $P_b$, that only differ by $\unit[0.2]{\mu C/m^2}$ from each other.

For a magnetic field along $\bi{b}$, we observe a weak systematic increase of the transition temperature $T_{\mathrm{C}}\simeq \unit[11.6]{K}$ and also the magnitude of the electric polarization is slightly enlarged, but its orientation remains almost unchanged. In contrast, magnetic fields along \mbox{$\bi{b}\!\!\times\!\!\bi{c}$} or $\bi{c}$ strongly change the electric polarization. For both field directions, the $ac$-plane components $P_{b\times c}$ and $P_c$ are suppressed at the respective spin-flop fields $B_{\mathrm{SF}}^{b\times c}\simeq \unit[5]{T}$ or $B_{\mathrm{SF}}^c\simeq \unit[2]{T}$. Moreover, for \mbox{$\bi{B}||\bi{b}\!\!\times\!\!\bi{c}$} also the component $P_b$ is suppressed in the high-field range, but the suppression of $P_b$ sets in at slightly larger magnetic fields, compared to the suppression of the other two components $P_{b\times c}$ and $P_c$, and it is preceded by a slight increase of that component with a maximum at $B_{\mathrm{SF}}^{b\times c}$. For $\bi{B}||\bi{c}$, the component $P_b$ is first slightly suppressed in the vicinity of the corresponding spin-flop field, but then $P_b$ again increases with increasing field until $P_b\simeq \unit[3]{\mu C/m^{2}}$ is reached at \unit[15]{T}. This magnetic-field behaviour is illustrated in figure~\ref{figure2b}, where the components  $P_{b\times c}$, $P_b$ and $P_c$ are displayed as functions of the magnetic field for \mbox{$\bi{B}||\bi{b}\!\!\times\!\!\bi{c}$} and $\bi{B}||\bi{c}$ at a constant temperature of \unit[3.5]{K}. The data are taken from the temperature-dependent polarization measurements shown in figure~\ref{figure2}. Due to the anomalous magnetic-field dependence of the component $P_b$, described above, which is different from that of the components within the $ac$ plane, one may speculate that a different underlying mechanism could be responsible for its generation. 

The magnetic-field dependent modifications of the electric polarization of  NaFeGe$_2$O$_6$ can be summarized as follows: In zero magnetic field the spontaneous polarization is mainly lying within the $ac$ plane with a small component along $\bi{b}$. A magnetic field along $\bi{b}$, does not cause any reorientation of the electric polarization.  
 A magnetic field along \mbox{$\bi{b}\!\!\times\!\!\bi{c}$} or $\bi{c}$ causes a strong suppression of the electric polarization within the $ac$ plane, which coincides with the respective spin-flop fields at $\simeq \unit[5]{T}$ and $\simeq \unit[2]{T}$, respectively. For \mbox{$\bi{B}||\bi{b}\!\!\times\!\!\bi{c}$} also the component $P_b$ is suppressed, but this suppression sets in above $B_{\mathrm{SF}}^{b\times c}$ while close to $B_{\mathrm{SF}}^{b\times c}$, $P_b$ shows a weak maximum.  For $\bi{B}||\bi{c}$, $P_b$ has a minimum around $B_{\mathrm{SF}}^{c}$ and is then even slightly increased with field. In both cases the total electric polarization is strongly decreased.

\begin{figure}[t]
\includegraphics[width=\textwidth]{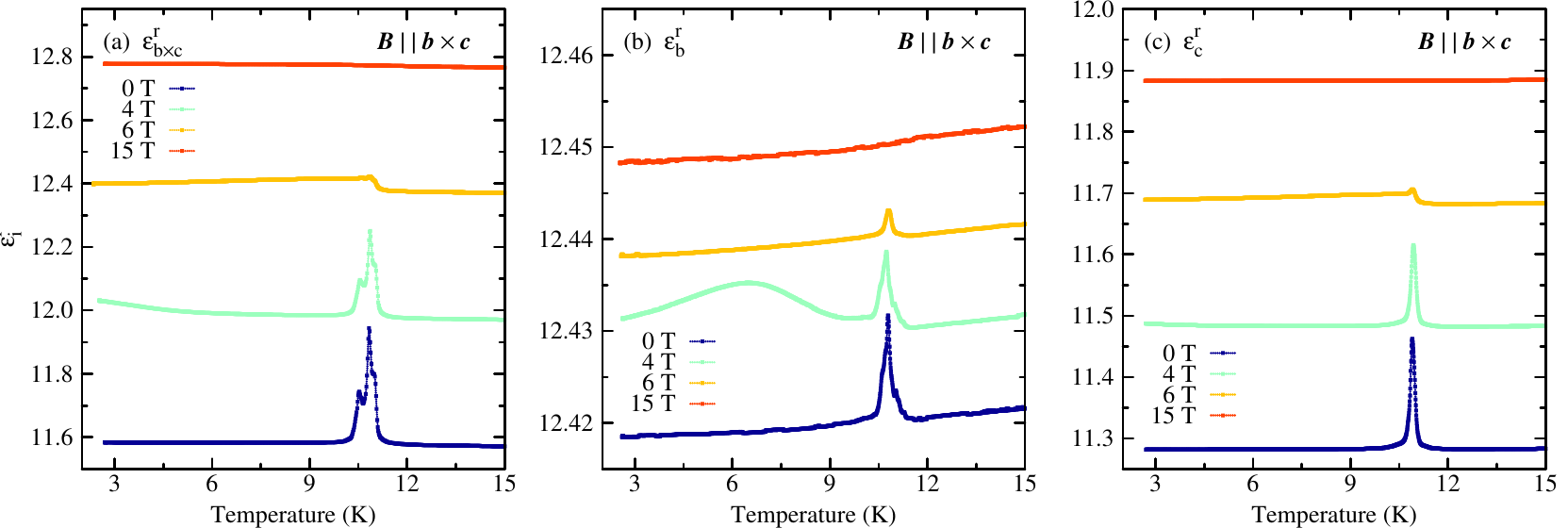}
\caption{Longitudinal components of the dielectric tensor $\epsilon^{r}_{i}$ ($i=b \times c, b, c$) of NaFeGe$_2$O$_6$ as functions of temperature for \mbox{$\bi{B}||\bi{b}\!\!\times\!\!\bi{c}$}. With increasing field, the curves are shifted with respect to each other by constant offsets of 0.4 in (a), 0.01 in (b) and 0.2 in (c).}
\label{figure3}
\end{figure}

The ferroelectric ordering is also reflected by distinct anomalies in the temperature dependences of the corresponding longitudinal components $\epsilon_{i}^{r}$ of the dielectric tensor. This is illustrated in figure~\ref{figure3} for representative measurements of the temperature dependences of $\epsilon_{i}^{r}$ ($i=b\times c, b, c$) for \mbox{$\bi{B}||\bi{b}\!\!\times\!\!\bi{c}$}. Below the corresponding spin-flop transition, all components $\epsilon_{i}^{r}$ display spiky anomalies for temperatures between \unit[9]{K} and \unit[12]{K}.  Above $B_{\mathrm{SF}}\simeq \unit[5]{T}$, the anomalies of all components $\epsilon_{i}^{r}$ essentially vanish. In addition, the anomalies of $\epsilon_{b \times c}^{r}$ have two maxima. An explanation of this behaviour is, however, still missing. Combining the observed dielectric constants with the measured electric polarization allows to conclude that the spikes signal the transitions to ferroelectric phases in which the spontaneous polarization either lies mainly in the $ac$ plane or is oriented nearly along the $\bi{b}$ axis.

\begin{figure}[t]
\includegraphics[width=\textwidth]{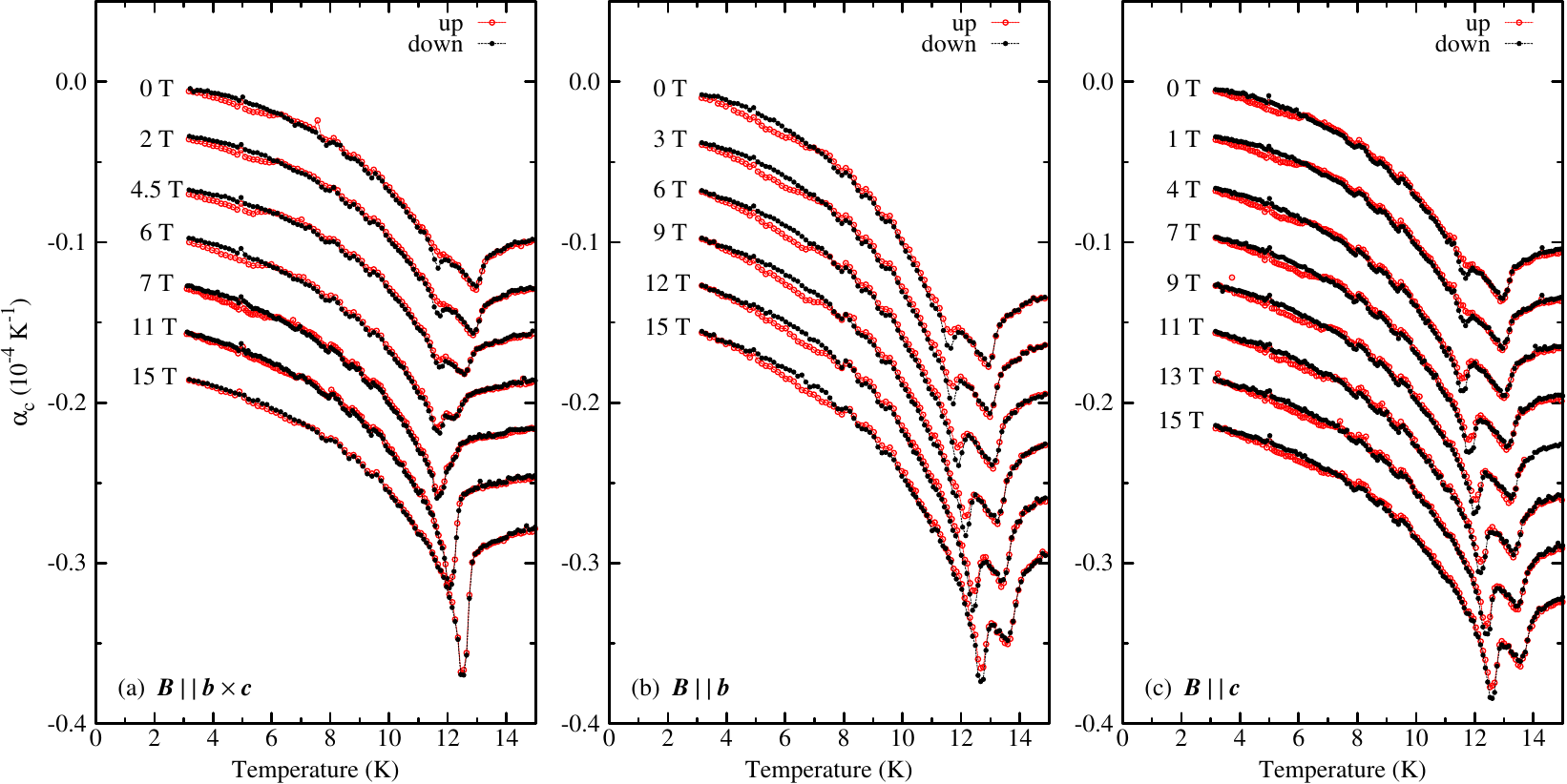}
\caption{Thermal expansion of NaFeGe$_2$O$_6$ along the $\bi{c}$ direction for magnetic fields applied either parallel to the \mbox{$\bi{b}\!\!\times\!\!\bi{c}$}, $\bi{b}$ or $\bi{c}$ axis (from left to right). For clarity, with increasing field strengths the curves are shifted with respect to each other by constant offsets of $\unit[-0.03\cdot 10^{-4}]{K^{-1}}$. Measurements with increasing and decreasing temperature are plotted as red and black symbols, respectively.}
\label{figure4}
\end{figure}

In order to determine the temperature versus magnetic field phase diagrams of NaFeGe$_2$O$_6$ thermal-expansion and magnetostriction measurements along $\bi{c}$ for magnetic fields applied parallel to the \mbox{$\bi{b}\!\!\times\!\!\bi{c}$}, $\bi{b}$ or $\bi{c}$ axis were performed. The thermal-expansion coefficient $\alpha_{c}(T)=1/L_c^0\cdot \partial \Delta L_c/\partial T$ measured as a function of increasing or decreasing temperature at constant magnetic fields is presented in figure~\ref{figure4}. Magnetic fields applied either along $\bi{b}$ or $\bi{c}$ have only little influence. In both cases, the zero-field transition temperatures $T_{\rm N} \simeq \unit[13]{K}$ and $T_{\rm C} \simeq \unit[11.6]{K}$ continuously increase with increasing field. Moreover, the anomaly at $T_{\rm C} \simeq \unit[11.6]{K}$ grows in intensity with increasing magnetic fields, compared to the anomaly at $T_{\rm N} \simeq \unit[13]{K}$. As can be seen in figure~\ref{figure4}\,(a), for \mbox{$\bi{B}||\bi{b}\!\!\times\!\!\bi{c}$} the two transitions occurring at $T_{\rm N}$ and $T_{\rm C}$ in zero field converge with increasing field strength until above \unit[7]{T} only one single transition can be resolved. The corresponding anomaly gets stronger with further increasing magnetic field.

\begin{figure}[t]
\includegraphics[width=\textwidth]{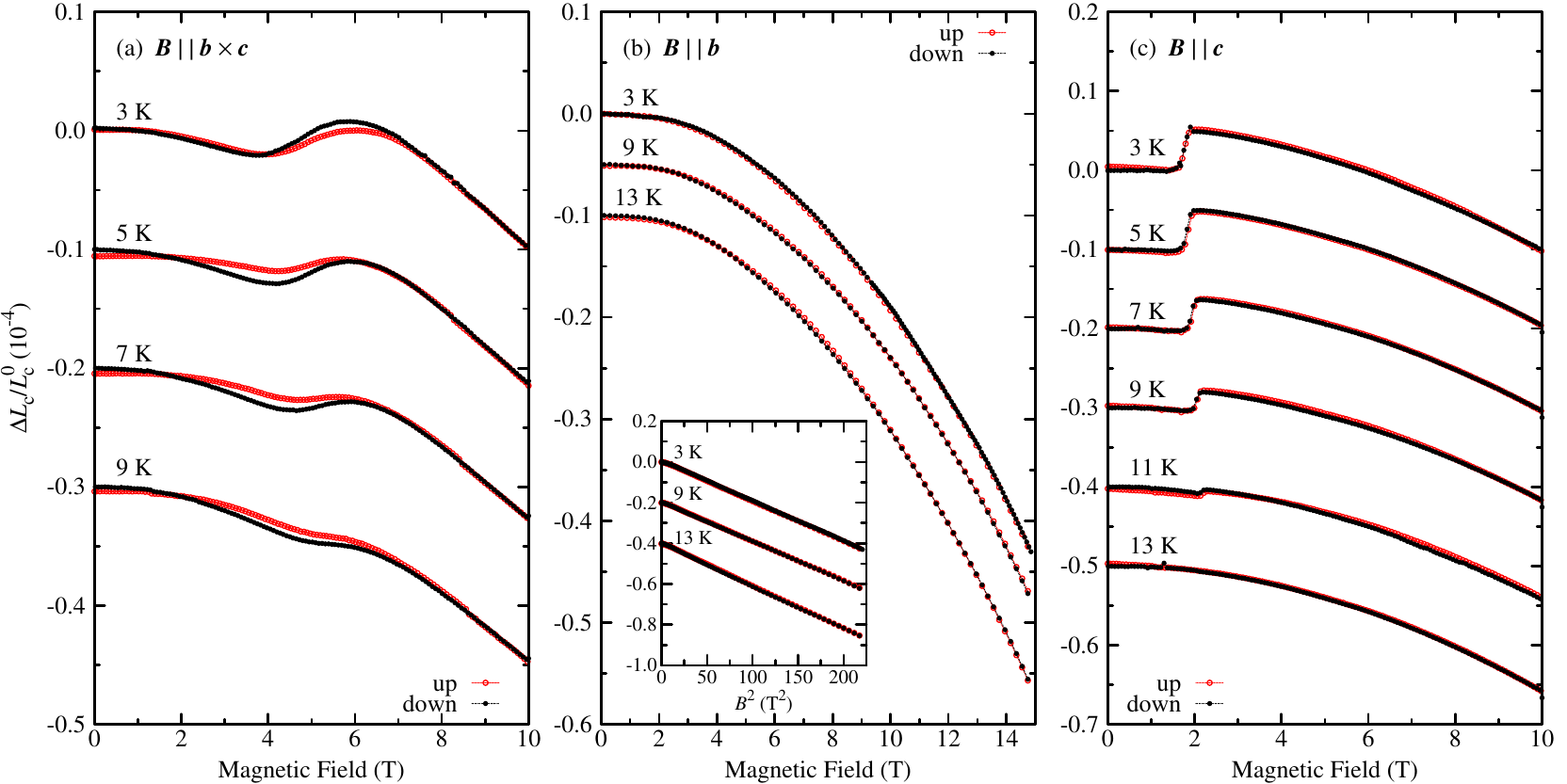}
\caption{Magnetostriction of the $\bi{c}$ axis of  NaFeGe$_2$O$_6$ for magnetic fields parallel to  \mbox{$\bi{b}\!\!\times\!\!\bi{c}$}, $\bi{b}$ or $\bi{c}$. For clarity,  with increasing temperature the curves are shifted with respect to each other by constant offsets of $-0.1\cdot 10^{-4}$ in (a) and (c). In the main panel of (b) an offset of $-0.05\cdot 10^{-4}$ was used and the inset shows the same data versus $B^2$ with an offset of $-0.2\cdot 10^{-4}$. Measurements with increasing and decreasing magnetic field are plotted as red and black symbols, respectively.}
\label{figure5}
\end{figure}

Representative magnetostriction measurements along the $\bi{c}$ direction for magnetic fields applied either parallel to \mbox{$\bi{b}\!\!\times\!\!\bi{c}$}, $\bi{b}$ or $\bi{c}$ are displayed in Fig.~\ref{figure5}.  Here, the relative length changes $\Delta L_c(T_0,\bi{B})/L_c^0$ as a function of the magnetic field \mbox{$\bi{B}||\bi{b}\!\!\times\!\!\bi{c}$}, $\bi{B}||\bi{b}$ or $\bi{B}||\bi{c}$ are shown. The data have been studied up to a maximum field of \unit[15]{T}, but for clarity the field scales of figure \ref{figure5}\,(a) and (c) have been limited to \unit[10]{T}, because there are no  anomalies in the higher field range. For $\bi{B}|| \bi{b}$, a quadratic magnetostriction $\Delta L_{c}/L_{c}^0\propto B^2$ is observed over the entire field range, see the inset of figure~\ref{figure5}\,(b), as it is typical for materials with a linear field dependece of the magnetization. For $\bi{B}|| \bi{c}$, the corresponding spin-flop transition around $B_{\mathrm{SF}}\simeq \unit[2]{T}$ coincides with positive, almost discontinuous length changes of $\Delta L_{c}/L_{c}^0$, indicating that the spin-flop transition is of first order. With increasing temperature, the spin-flop transition shifts towards higher magnetic-field strength and the length changes decrease.
For \mbox{$\bi{B}||\bi{b}\!\!\times\!\!\bi{c}$}, at \unit[3]{K} there is a phase transition with a blurred, positive length change $\Delta L_{c}/L_{c}^0$, which again coincides with the corresponding spin-flop transition at $B_{\mathrm{SF}}\simeq \unit[5]{T}$, see figures~\ref{figure5}\,(a) and~\ref{figure1}. Already the magnetization, as well as the electric polarization data indicated that the spin-flop transition for this field direction is rather broad. Moreover, there is a slight hysteresis between the measurements with increasing and decreasing magnetic field. As a function of increasing temperature the transition shifts towards higher magnetic fields and the corresponding length changes decrease.

\section{Phase diagrams and conclusion}

\begin{figure}[t]
\includegraphics[width=\textwidth]{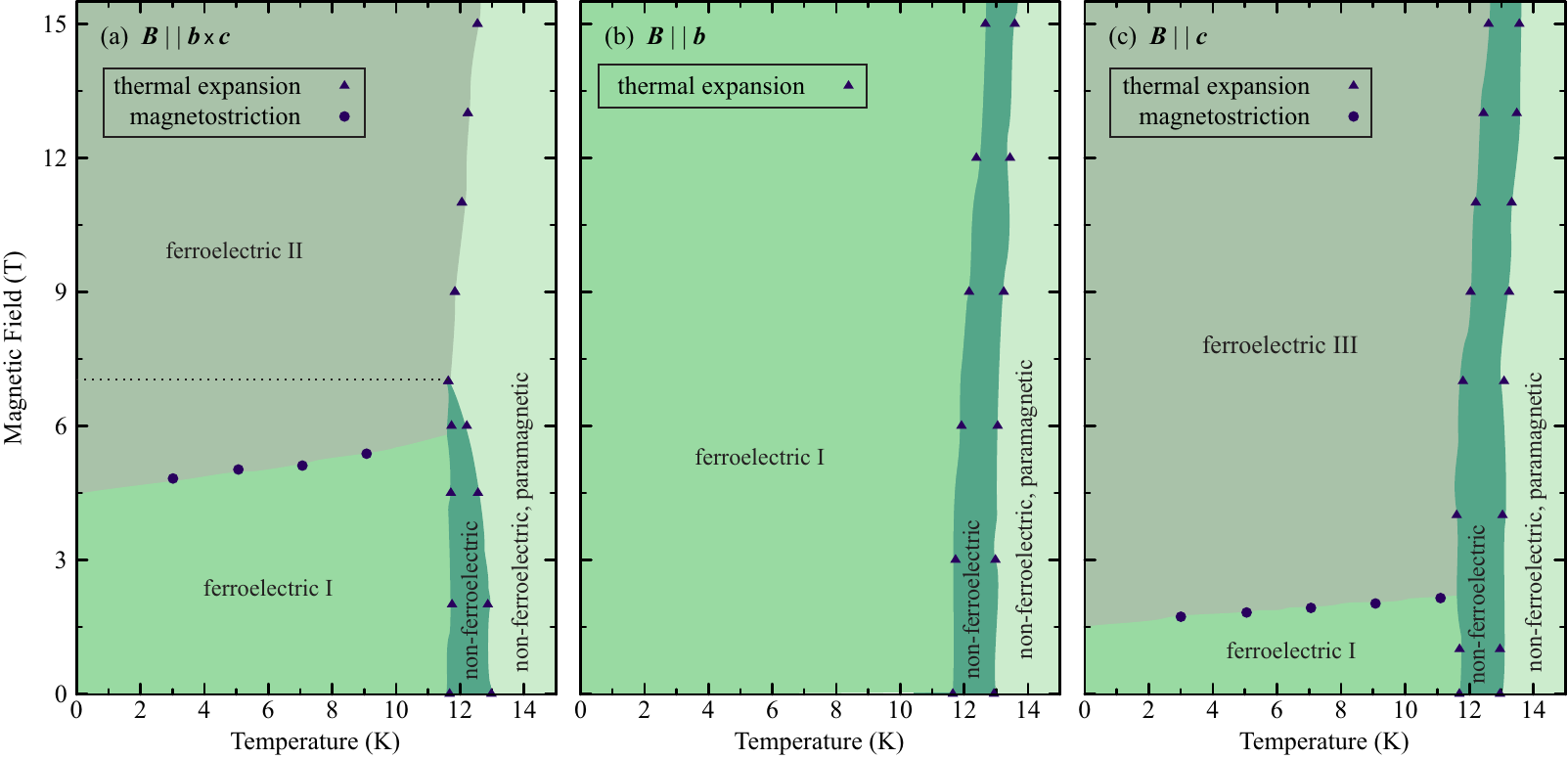}
\caption{Magnetic field versus temperature phase diagrams of NaFeGe$_2$O$_6$ for $\bi{B}$ parallel to \mbox{$\bi{b}\!\!\times\!\!\bi{c}$}, $\bi{b}$ or $\bi{c}$. The phase boundaries are based on the thermal-expansion ($\alpha_c$) and magnetostriction ($\lambda_c$) measurements.}
\label{figure6}
\end{figure}

By combining all the results of the present work, detailed magnetic field versus temperature phase diagrams are derived for $\bi{B}$ parallel to \mbox{$\bi{b}\!\!\times\!\!\bi{c}$}, $\bi{b}$ and $\bi{c}$, see figure~\ref{figure6}. The critical fields and temperatures are determined from the anomalies in $\Delta L_{c}(T,\bi{B})/L_{c}^0$ from the thermal-expansion and magnetostriction measurements. For most of the detected phase transitions there are also anomalies in the polarization and magnetization measurements and within the experimental uncertainties their positions agree with the anomalies of $\Delta L_{c}(T,\bi{B})/L_{c}^0$. In addition, the electric polarization and magnetization data also allow to identify the dielectric and magnetic properties of the various phases. In zero magnetic field, NaFeGe$_2$O$_6$ undergoes a phase transition at $T_{\rm N}\simeq$\,$\unit[13]{K}$ from its paramagnetic and non-ferroelectric high-temperature phase to an antiferromagnetically ordered, non-ferroelectric phase. The magnetic-susceptibility measurements indicate a collinear spin structure with the spins being oriented mainly along the \mbox{$\bi{b}\!\!\times\!\!\bi{c}$} axis in this phase, see figure~\ref{figure1}.  On further cooling, a spin reorientation occurs at $T_{\rm C}\simeq$\,$\unit[11.6]{K}$ leading to an $XY$ anisotropy and to the onset of ferroelectricity. The ferroelectric phase extends down to the experimental low-temperature limit of \unit[2.5]{K}. The spontaneous polarization of this  ferroelectric phase I has an absolute value of $\simeq \unit[32]{\mu C/m^{2}}$ and is  mainly lying within the $ac$ plane with a small component \mbox{along $\bi{b}$.}

A magnetic field $\bi{B}||\bi{b}$ is perpendicular to the magnetic easy plane and has only little influence. As shown in figure~\ref{figure6}\,(b), it only causes a weak simultaneous increase of both transition temperatures $T_{\rm C}(\bi{B})$ and $T_{\rm N}(\bi{B})$. Moreover, the magnetization linearly increases with field there is also a weak increase of the electric polarization, see figures~\ref{figure1} and~\ref{figure2}.

Magnetic fields along \mbox{$\bi{b}\!\!\times\!\!\bi{c}$} or $\bi{c}$ induce spin-flop transitions, which are accompanied by strong modifications of the electric polarization. With increasing $\bi{B}||\bi{b}\!\!\times\!\!\bi{c}$, above $B_{\mathrm{SF}}^{b\times c}\simeq \unit[5]{T}$ the electric polarization within the $ac$ plane is continuously suppressed. Interestingly, the suppression of $P_b$ sets in at a slightly higher field strength near \unit[7]{T} and is preceded by a maximum at $B_{\mathrm{SF}}^{b\times c}\simeq \unit[5]{T}$. Around \unit[7]{T} also the intermediate antiferromagnetically ordered but non-ferroelectric phase vanishes, which is illustrated by the horizontal dashed line in figure~\ref{figure6}\,(a). For $\bi{B}||\bi{c}$, only the electric polarization within the $ac$ plane is continuously suppressed above $B_{\mathrm{SF}}^c\simeq \unit[2]{T}$, while the polarization along $\bi{b}$ has a minimum at $B_{\mathrm{SF}}$ and then grows with increasing field strength. The intermediate antiferromagnetially ordered, non-ferroelectric phase stays present for the whole investigated magnetic-field range in this case.

Compared to aegirine (of the composition Na$_{1.04}$Fe$_{0.83}$Ca$_{0.04}$Mn$_{0.02}$Al$_{0.01}$Ti$_{0.08}$Si$_2$O$_6$), the second multiferroic compound among the pyroxenes, there are some clear differences in NaFeGe$_2$O$_6$. First of all, the spin structure of the multiferroic phase of aegirine forms a spiral with the spins lying mainly within the $ac$ plane and a propagation vector along the monoclinic $\bi{b}$ axis~\cite{Baum2013}. In contrast, in the multiferroic phase of NaFeGe$_2$O$_6$ the spins apparently form a cycloid within the $ac$ plane~\cite{Drokina2010, Redhammer2011, Drokina2011}. Secondly, the presence of all three components of the electric polarization in NaFeGe$_2$O$_6$ indicate a triclinic symmetry $1$ for its multiferroic phase. The spontaneous electric polarization in the multiferroic phase of aegirine points along the monoclinic $\bi{b}$ axis and is about a factor of 2.5 smaller than the  electric polarization of NaFeGe$_2$O$_6$~\cite{Jodlauk2007}. Consequently the symmetry of the multiferroic phase of aegirine is higher compared to that of NaFeGe$_2$O$_6$ with the point group $2$. Finally, also the magnetic-field dependence of the electric polarization is different in both compounds. In aegirine a magnetic field within the $ac$ plane causes a rotation of the electric polarization from parallel $\bi{b}$ towards $\bi{c}$, which is connected with a strong decrease~\cite{Jodlauk2007}. In contrast, in NaFeGe$_2$O$_6$ a magnetic field within the $ac$ plane causes a strong suppression of the electric polarization within this plane. For $\bi{B}||\bi{b}\!\!\times\!\!\bi{c}$ also $P_b$ is suppressed, while for $\bi{B}||\bi{c}$ this component is even slightly increased.

The spontaneous electric polarization of many spin-driven multiferroics with cycloidal spin structures can be described by the relation $\bi{P}\propto\bi{e_{ij}}\times(\bi{S^{(i)}}\times\bi{S^{(j)}})$~\cite{Sergienko2006, Katsura2005}. Here, however, the prediction of the relation is inconsistent with the present results of  NaFeGe$_2$O$_6$, if a cycloid within the $ac$ plane is assumed. The zero-field orientation of the electric polarization, found in this work, indicates  the presence of a more complex spin structure than that reported in~\cite{Drokina2010, Redhammer2011, Drokina2011} with a finite spin component along $\bi{b}$ within the ferroelectric phase I. Furthermore, the different magnetic-field characteristics of $P_b$ compared to the $ac$-plane components indicate a possibly different underlying mechanism for the generation of $P_b$. In order to clarify the microscopic mechanisms leading to multiferroicity in NaFeGe$_2$O$_6$ and in order to resolve the inconsistencies and open questions discussed above, more detailed information about the magnetic structure is needed. Therefore, as a future task neutron-diffraction experiments should be performed on single crystals of NaFeGe$_2$O$_6$.

\section*{Acknowledgements}
This work was supported by the Deutsche Forschungsgemeinschaft via SFB 608 and through the Institutional Strategy of the University of Cologne within the German Excellence Initiative.

\end{document}